\documentclass[twocolumn,11pt]{article}

\usepackage[T2A]{fontenc}
\usepackage[cp1251]{inputenc}
\usepackage{amsmath}
\usepackage{amssymb}
\usepackage{textcase}
\usepackage{bm}
\usepackage{multicol}
\usepackage{tabularx}
\usepackage{euscript}
\usepackage{calc}
\usepackage{cite}
\usepackage{indentfirst}
\usepackage{mathtext}
\usepackage{titlesec}
\usepackage{graphicx}
\usepackage[a4paper, left=1.75cm, right=1.5cm, top=1.9cm, bottom=2.6cm]{geometry}

\titleformat*{\section}{\centering\large\uppercase}

\begin{document}

\twocolumn[
\begin{@twocolumnfalse}
\vspace{3em}
\centering
\begin{minipage}{0.84\textwidth}
\begin{center}
{\Large\uppercase{\bf 
Modified Monte Carlo method with the heat bath algorithm for a model cuprate
}}

\vspace{1em}
\textbf{\large 
Yu. D. Panov{*}, 
V. A. Ulitko{*}, 
D. N. Yasinskaya{*}, 
A. S. Moskvin{*\textsuperscript{,}**}
}
	
\vspace{1em}
	
{\small
{*}\textit{Ural Federal University, 620002 Ekaterinburg, 19 Mira street, Russia}

{**}\textit{M.N. Miheev Institute of Metal Physics of UD of RAS, 620108 Ekaterinburg, S. Kovalevskaya, 18, Russia}

}

\end{center}

{\small
The results of numerical simulation using a modified Monte Carlo method with a heat bath algorithm for the pseudospin model of cuprates are presented. The temperature phase diagrams are constructed for various degrees of doping and for various parameters of the model, and the effect of local correlations on the critical temperatures of the model cuprate is investigated. It is shown that, in qualitative agreement with the results of the mean field, the heat bath algorithm leads to a significant decrease in the estimate of critical temperatures due to more complete accounting of fluctuations, and also makes it possible to detect phase inhomogeneous states. The possibility of using machine learning to accelerate the heat bath algorithm is discussed.

\vspace{1em}
}
\vspace{2em}

\end{minipage}

\end{@twocolumnfalse}
]

	
\section*{Introduction}

The relationship between superconductivity and other competing order parameters remains a hotly debated topic in the physics of high-temperature superconducting cuprates  \cite{Keimer2015}. 
The versatility of the properties of these compounds makes it relevant to use effective models to describe them. 
Cuprates belong to a large family of Jahn-Teller magnets \cite{Moskvin2023}, in which the removal of orbital degeneracy occurs due to anti-Jahn-Teller disproportionation.  
Within the framework of the pseudospin model of cuprates  \cite{Moskvin2011,Moskvin2013} the different valence states of CuO$_4$ clusters in the  CuO$_2$ plane are described as components of the $S=1$ triplet.
Earlier, in the framework of the mean-field approximation, we obtained equations for the critical temperatures of various ordered phases for the pseudospin model of cuprates  \cite{Panov2019} and found the temperature phase diagrams in the approximation of pure phases  \cite{Moskvin2020} and taking into account phase separation \cite{Moskvin2021,Moskvin2022}. 

In this paper, we discuss the possibility and specific features of applying the heat bath algorithm of the Monte Carlo (MC) method to a pseudospin cuprate model with a local basis from the quartet of states of the CuO$_{4}$ cluster. 
First, the model used is briefly formulated and some basic results of the mean-field approximation are given. 
In particular, the effect of local charge correlations on the behavior of pure phases with only one order parameter is different from zero is investigated. 
The following describes the modified heat bath algorithm and discusses the possibility of using machine learning to speed it up. 
The results of numerical simulations are compared with the mean-field approximation, and the possibility of detecting phase inhomogeneous states is shown.

\section*{Mean field approximation}

In the framework of the effective pseudospin model of cuprates, the basis of the local Hilbert space of the CuO$_4$ cluster in the CuO$_2$ plane is chosen as a quartet of states $\left| 1 M s m \right\rangle$. 
The projections $M = {-}1,\, 0,\, {+}1$ of pseudospin $S=1$ correspond to the three multi-electron states of the cluster [CuO]$_4^{7-,6-,5-}$ (nominally Cu$^{1+, 2+, 3+}$). 
The states with $M = \pm1$ are the spin singlets, $s=0$, the state with $M = 0$ is a spin doublet, $s=1/2$. 

The Hamiltonian of the model cuprate can be written as follows:
\begin{equation}
\hat{H} 
	= \hat{H}_{ch} 
	+ \hat{H}_{exc} 
	+ \hat{H}_{tr}^{(1)} 
	+ \hat{H}_{tr}^{(2)} 
	- \mu \sum_i \hat{S}_{z,i} .
\label{eq:Ham0}
\end{equation}
The first term 
\begin{equation}
	\hat{H}_{ch} =
	\Delta \sum_i \hat{S}_{z,i}^2 
	+ V \sum_{\left\langle ij \right\rangle} \hat{S}_{z,i} \hat{S}_{z,j} 
\end{equation}
describes local ($\Delta=U/2$, where $U$ is the correlation parameter) and nonlocal ($V>0$) charge-charge correlations. 
Here ${\hat{S}}_z$  is the $z$-projection of the pseudospin $S=1$ operator. 
The summation is performed on $N$ sites of the square lattice, 
$\left\langle i j\right\rangle$ denotes nearest neighbors. 
The operator
\begin{equation}
	\hat{H}_{exc} =
	Js^2 \sum_{\langle ij \rangle} \hat{\boldsymbol{\sigma}}_i \hat{\boldsymbol{\sigma}}_j 
\end{equation}
describes the antiferromagnetic ($J>0$) Heisenberg exchange interaction for [CuO]$_4^{6-}$ centers, where the operators $\hat{\boldsymbol{\sigma}}=\hat{P}_0 \, \hat{\mathbf{s}}/s$ take into account the spin density at the site, $\hat{P}_0 = 1-\hat{S}_z^2$ is the spin $s=1/2$ operator \cite{Panov2016}.  
The operator
\begin{multline}
	\hat{H}_{tr}^{(1)}
	\;=\;
	- \sum_{\left\langle ij\right\rangle m} 
	\Big[  
	t_p \hat{P}_{m,i}^{{+}} \hat{P}_{m,j}^{}  +  t_n \hat{N}_{m,i}^{{+}} \hat{N}_{m,j}^{} 
	\\
	{} + 
	\frac{t_{pn}}{2} \big( \hat{P}_{m,i}^{{+}} \hat{N}_{m,j}^{} + \hat{N}_{m,i}^{{+}} \hat{P}_{m,j}^{} \big) 
	\Big]
	+ h.c. 
\end{multline}
describes three types of correlated single-particle transport \cite{Moskvin2011,Moskvin2013}: the terms proportional to $t_p$ ($t_n$) describe the transport of [CuO]$_4^{5-}$ ([CuO]$_4^{7-}$) centers on the lattice of [CuO]$_4^{6-}$ centers; 
the terms proportional to $t_{pn}$ are responsible for the disproportionation and recombination processes: $[\mathrm{CuO}]_4^{6-}+[\mathrm{CuO}]_4^{6-} \leftrightarrow [\mathrm{CuO}]_4^{7-}+[\mathrm{CuO}]_4^{5-}$. 
The spin index $m=\pm1/2$ corresponds to the projections of spin $s=1/2$, the orbital part of the operators $\hat{P}$ and $\hat{N}$ is expressed through the pseudospin $S=1$ operators: $\hat{P}^{+} = \big(\hat{S}_{+} + \hat{T}_{+}\big)/2$, 
$\hat{N}^{+} = \big(\hat{S}_{+} - \hat{T}_{+}\big)/2$, 
$\hat{T}_{+} = \hat{S}_{z} \hat{S}_{+} + \hat{S}_{+} \hat{S}_{z}$, 
$\hat{S}_{+} = \big( \hat{S}_{x} + i \hat{S}_{y} \big) / \sqrt{2}$. 
The operator
\begin{equation}
	\hat{H}_{tr}^{(2)}
	=	- t_b \sum_{\left\langle ij\right\rangle} 
	\big( \hat{S}_{{+}i}^2 \hat{S}_{{-}j}^2 + \hat{S}_{{+}j}^2 \hat{S}_{{-}i}^2 \big)
\end{equation}
describes two-particle transport (the local composite boson transport) \cite{Moskvin2011,Moskvin2013}): $[\mathrm{CuO}]_4^{5-}+[\mathrm{CuO}]_4^{7-} \leftrightarrow [\mathrm{CuO}]_4^{7-}+[\mathrm{CuO}]_4^{5-}$. 
The last sum with the chemical potential $\mu$ allows to take into account the charge constancy condition, $n = \left\langle \sum_i \hat{S}_{z,i} \right\rangle / N = const$.

In the mean-field approximation in \cite{Panov2019}, the equations of critical temperatures of ordered phases with only one of the possible order parameters is not zero were found. 
We introduce two sublattices A and B forming a staggered order on the square lattice and denote $C_{\alpha} = \big\langle \hat{C}_{i \in \alpha} \big\rangle$, $\alpha=A,B$. 

For the charge-ordered (CO) phase with the order parameter $L=\left(S_{z,A}-S_{z,B}\right)/2$ the equation for the critical temperature $T_{CO}$ has the form
\begin{equation}
	T = \frac{4 V \left( 1 - n^2 \right) \phi(n,T)}{1 + \phi(n,T)} , 
	\label{eq:TCO}
\end{equation}
where 
\begin{equation}
	\phi(n,T) = \sqrt{ \left( 1 - n^2 \right) e^{-2\Delta / T} + n^2 } . 
\end{equation}
This equation generalizes the well-known result for the critical temperature of charge ordering in the hard-core bosons model \cite{Micnas1990}.

For the antiferromagnetic (AFM) phase, the order parameter is defined by the expression $\boldsymbol{l} = \big( \boldsymbol{\sigma}_{A} - \boldsymbol{\sigma}_{B} ) / 2$. 
The critical temperature $T_{AFM}$ can be found from the equation 
\begin{equation}
	T = \frac{4 J s^{2} \left( 1 - n^2 \right) }{1 + \phi(n,T)} . 
	\label{eq:TAFM}
\end{equation}

The phase with a nonzero average $\big\langle \hat{S}_{+}^2 \big\rangle$ by analogy with the model of local bosons \cite{Micnas1990} can be called a bose superfluid (BS). 
Equation for the critical temperature $T_{BS}$
\begin{equation}
	T = 4 t_{b} n \left[ \ln \frac{\left(1 + n\right)\left(\phi(n,T) + n\right)}{\left(1 - n\right)\left(\phi(n,T) - n\right)} \right]^{-1} 
	\label{eq:TBS}
\end{equation}
generalizes the known result \cite{Micnas1990} and leads to an expression for $T_{BS}$ in the model of local bosons at $\Delta\to -\infty$. 

By analogy, for phases with non-zero order parameters 
$\big\langle \hat{P}_{m}^{+} \big\rangle$ and $\big\langle \hat{N}_{m}^{+} \big\rangle$  we can find in the case $t_{pn}=0$ the equations for critical temperatures $T_p$:
\begin{equation}
	T = 2 t_{p} \frac{ \left(1+n\right)\left[1-2n-\phi(n,T)\right] }{ \left[ 1+\phi(n,T) \right] \ln \left( \frac{1 - n}{\phi(n,T) + n} \right) } , 
	\label{eq:Tp}
\end{equation} 
and $T_{n}$:
\begin{equation}
	T = 2 t_{n} \frac{ \left(1-n\right)\left[1+2n-\phi(n,T)\right] }{ \left[ 1+\phi(n,T) \right] \ln \left( \frac{1 + n}{\phi(n,T) - n} \right) } . 
	\label{eq:Tn}
\end{equation} 
In these phases, correlated single-particle transport of hole (P) or electron (N) type is realized.

\begin{figure*}
	\centering
	\includegraphics[width=0.8\linewidth]{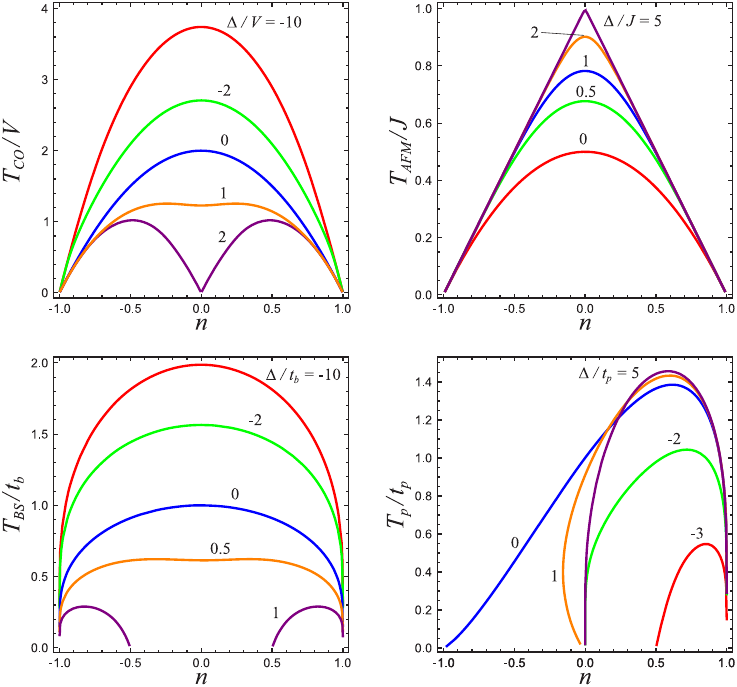}
	\caption{The concentration dependences of critical temperatures for different values of the parameter of local charge-charge correlations.}
	\label{fig:tc-mfa}
\end{figure*}

Fig.~\ref{fig:tc-mfa} shows the concentration dependences of critical temperatures obtained from equations (\ref{eq:TCO}--\ref{eq:Tp}) for different values of the parameter $\Delta$.

The critical temperatures of both the CO and BS phases have the highest values at $\Delta\to-\infty$ and reach a maximum at n=0. 
As $\Delta$ increases, the values of $T_{CO}$ and $T_{BS}$ decrease, and at positive values of the parameter of local charge-charge correlations CO and BS phases become unstable at $n=0$.
The concentration dependence of $T_{CO}$ at $\Delta / V \geq 2$ tends to the limiting value $T_{CO}=4V|n|\left(1-|n|\right)$, and there is no BS phase at $\Delta / t_{b} \geq 2$. 

The phases AFM and P are not realized at large negative $\Delta$, but exist at $\Delta\rightarrow+\infty$. 
At $\Delta\geq0$, the concentration dependence of $T_{AFM}$ has a maximum at $n=0$, changing from $T_{AFM}=2Js^2\left(1-n^2\right)$ at $\Delta=0$ to $T_{AFM}=4Js^2\left(1-\left|n\right|\right)$ at $\Delta\rightarrow+\infty$. 
For the P phase, the solutions of equation (10) appear at $\Delta/t_p>-4$ and at $\Delta\rightarrow+\infty$ tend to some limiting dependence $T_p\left(n\right)$ in the region $n>0$. 
For the critical temperature of the N phase, the relation $T_n\left(n\right)=T_p\left(-n\right)$ is valid, provided that $t_p$ is replaced by $t_n$.

The question of the stability of ordered phases at intermediate values of the parameter of local charge-charge correlations leads to the cumbersome problem of analyzing the system of nonlinear equations for the order parameters \cite{Panov2019}. 
This problem can be considered using numerical modeling methods when the heat bath algorithm is employed.

\section*{The heat bath algorithm}

For numerical modeling, we use the heat bath algorithm of the MC method \cite{Miyatake1986}. 
Now, let's briefly formulate it for model \eqref{eq:Ham0}. 
We can write the wave function of the system as a product of wave functions at each site:
\begin{equation}
	\left| \Psi \right\rangle = \prod_{\nu} \left| \psi_{\nu} \right\rangle , \;\;
	\left| \psi_{\nu} \right\rangle 
	= \sum_{Mm} a_{Mm}^{\nu} \left| 1Msm \right\rangle .
	\label{eq:Psi}
\end{equation}
We construct the Hamiltonian of node $\nu$ by averaging over the states of all other sites $\mu\neq\nu$:
\begin{equation}
	\hat{H}_{\nu} = \big\langle \tilde{ \Psi }_{\nu} \big| \hat{H} \big| \tilde{ \Psi }_{\nu} \big\rangle , \quad
	\big| \tilde{ \Psi }_{\nu} \big\rangle = \prod_{\gamma \neq \nu} \left| \psi_{\gamma} \right\rangle .
\end{equation}
Taking into account the form of the Hamiltonian \eqref{eq:Ham0}, we obtain
\begin{multline}
	\hat{H}_{\nu} = \Delta \hat{S}_{z}^{2} 
	+ \sum_{ \nu' } \left( 
	V \big\langle \hat{S}_{z} \big\rangle_{\nu'} \, \hat{S}_{z} 
  	+ J s^2 \big\langle \hat{\boldsymbol{\sigma}} \big\rangle_{\nu'} \,  \hat{\boldsymbol{\sigma}} 
  	\right) \\
	- \sum_{\nu', m} 
	\Big[  
	t_p \big\langle \hat{P}_{m}^{+} \big\rangle_{\nu'} \, \hat{P}_{m}^{}  
	+  t_n \big\langle \hat{N}_{m}^{+} \big\rangle_{\nu'} \, \hat{N}_{m}^{} 
	\\
	{} + 
	\frac{t_{pn}}{2} \big(  \big\langle \hat{P}_{m}^{+} \big\rangle_{\nu'} \, \hat{N}_{m}^{} 
	+ \big\langle \hat{N}_{m}^{{+}} \big\rangle_{\nu'} \, \hat{P}_{m}^{} \big) + h.c.
	\Big]
	- \mu \hat{S}_{z} .
	\label{eq:Hc}
\end{multline}
Here, the $\big\langle \hat{A} \big\rangle_{\nu'} = \big\langle \psi_{\nu'} \big| \hat{A} \big| \psi_{\nu'} \big\rangle$ values act as external fields acting on the site $\nu$ from the site $\nu'$, and, in accordance with the chosen approximation, summation over the nearest neighbors to the site $\nu$ is assumed. 

The change of the site state is realized by the heat bath algorithm. 
For this purpose, we solve the eigenvalue problem at site $\nu$: 
\begin{equation}
	\hat{H}_{\nu} \left| \psi_{\nu,k} \right\rangle 
	= \varepsilon_{k} \left| \psi_{\nu,k} \right\rangle , \quad k = 1, \ldots 4 ,
\end{equation}
and construct the distribution function:
\begin{equation}
	F(n) = \sum_{k=1}^{n} p_k , \quad
	p_{k} = \frac{e^{- \beta\varepsilon_{k} }}{\sum_{l} e^{- \beta\varepsilon_{l} }} .
\end{equation} 
From the inequality $F(n-1) < \xi \leqslant F(n)$, where $\xi$ is a random variable uniformly distributed on the unit interval, we find the number n and choose the function $\left| \psi_{\nu,n} \right\rangle$ as the new state for site $\nu$. 
Then, using the wave function \eqref{eq:Psi} for the current state of the system, we calculate the necessary quantum mechanical averages and collect statistics on the MC steps in the standard way.

The method of constructing the Hamiltonian ${\hat{H}}_\nu$ allows direct generalization to clusters of several sites and accounting for interactions and transport between more distant neighbors. 
The structure of ${\hat{H}}_\nu$ coincides with the Hamiltonian of an ideal system in the mean-field method based on Bogoliubov's inequality (in the case of a cluster of several lattice sites, with the Hamiltonian of a Bethe cluster), but the nonuniform character of molecular fields, which are determined by the current state of the system, is taken into account in the heat bath algorithm. 
In contrast to the classical version of the MC method based on the Metropolis algorithm and assuming continuous change of observables, the quantum nature of the state change in the elementary MC step in the thermostat algorithm removes the divergence of the specific heat at low temperatures. 
The limitations of the algorithm include the high labor consumption of the elementary MC step associated with the need to solve the eigenvalue problem for the Hamiltonian at the site.

We also considered the possibility of speeding up the algorithm using machine learning techniques. 
As a preliminary study, we solved the problem of recovering the eigenvalues $\varepsilon_k$ of the matrices of the Hamiltonian ${\hat{H}}_\nu$ without its direct diagonalization. 
Here we take into account that all matrices ${\hat{H}}_\nu$ are sparse and have the same structure. This allows to use not the whole matrix, but only its unique (nonrepeating) elements as input data when training the model. 
We used a neural network of 4 fully connected layers with 128, 64, 32, 16 neurons as the training model and the Adam algorithm as the optimizer. 
The application of the neural network allowed to speed up the process of obtaining the $\varepsilon_k$ eigenvalues by an order of magnitude; however, the procedure of restoring the wave function $\left|\psi_{\nu,k}\right\rangle$ is comparable to the procedure of diagonalization in terms of labor input. 
We believe that extending this approach to the $\left|\psi_{\nu,k}\right\rangle$ recovery procedure will significantly speed up the entire algorithm.

\section*{Numerical simulation results}

Modeling was performed on a square lattice of size $32\times32$ with the periodic boundary conditions. 
A parallel version of the program for GPU was implemented using CUDA SDK tools. 
The cuSOLVER library was used for matrix diagonalization. 

\begin{figure}
	\centering
	\includegraphics[width=1\linewidth]{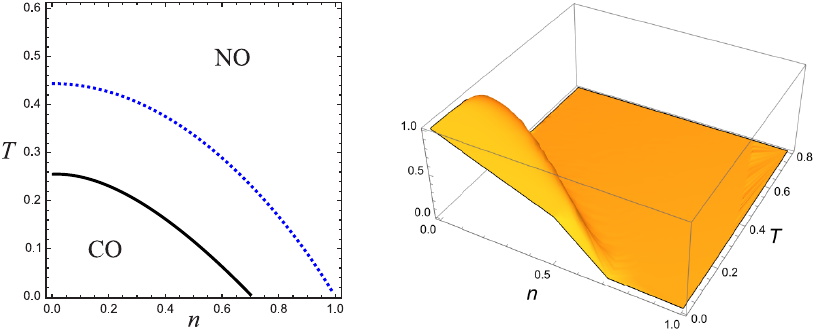}
	\caption{\small 
		Values of the critical temperature $T_{CO}$ obtained in the mean-field approximation (dotted line) and in the MC method (solid line). The values of the charge order parameter are shown on the right panel. The nonzero model parameters are $\Delta=0.1$, $V=0.25$.
		}
	\label{fig:co-phasediag}
\end{figure}

\begin{figure}
	\centering
	\includegraphics[width=1\linewidth]{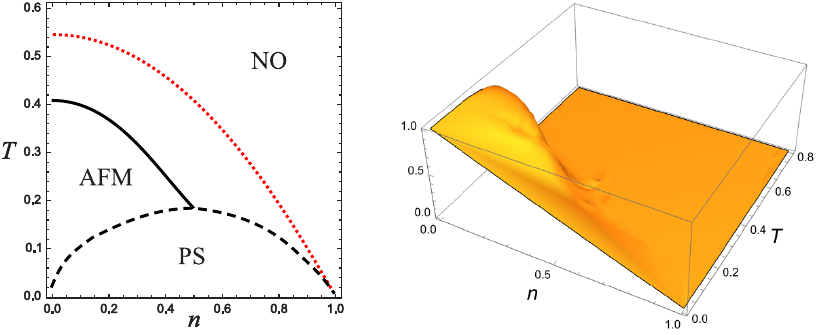}
	\caption{\small 
		Values of the critical temperature $T_{AFM}$ obtained in the mean-field approximation (dotted line) and in the MC method (solid line). The dashed line shows the boundary of the phase separation region. The values of the antiferromagnetic order parameter are shown on the right panel. The nonzero model parameters are $\Delta=0.1$, $J=1$.
		}
	\label{fig:afm-phasediag}
\end{figure}

\begin{figure}
	\centering
	\includegraphics[width=1\linewidth]{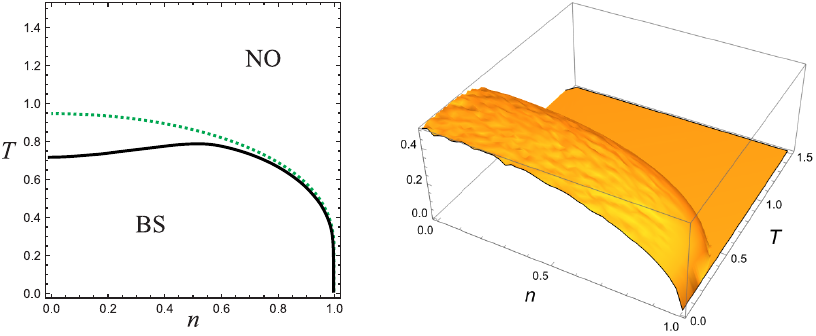}
	\caption{\small 
		Values of the critical temperature $T_{BS}$ obtained in the mean-field approximation (dotted line) and in the MC method (solid line). The values of the order parameter of the superfluid phase are shown on the right panel. The nonzero model parameters are $\Delta=0.1$, $t_b=1$.
		}
	\label{fig:bs-phasediag}
\end{figure}

\begin{figure}
	\centering
	\includegraphics[width=1\linewidth]{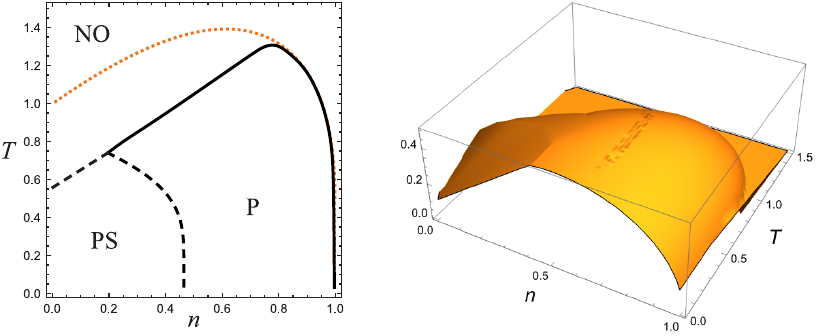}
	\caption{\small 
		Values of the critical temperature $T_p$ obtained in the mean-field approximation (dotted line) and in the MC method (solid line). The dashed line shows the boundary of the phase separation region. The values of the order parameter for the P phase are shown on the right panel. The nonzero model parameters are $\Delta=0.1$, $t_p=1$.
		}
	\label{fig:fl-phasediag}
\end{figure}

The results of numerical simulations and comparisons with analytical results in the mean-field approximation are shown in Figs.~\ref{fig:co-phasediag} -- \ref{fig:fl-phasediag}.
To obtain $T-n$ diagrams, calculations were performed for a sufficiently wide range of chemical potential values. 
At each $\left(T,\mu\right)$ point, the calculation started from a new random state of the system, and, after the annealing procedure, the average values of $n$ and all necessary thermodynamic parameters were calculated.

\bigskip
\bigskip

The values of the critical temperature of charge ordering $T_{CO}$ are shown in Fig.~\ref{fig:co-phasediag}. 
The dotted line shows the solution of equation \eqref{eq:TCO}, the solid line corresponds to the level line of the charge order parameter at the value of 0.01, the region of smaller values of the order parameter is designated as the disordered (NO) phase. 
Taking into account fluctuations of molecular fields in the numerical simulation leads to the expected decrease of the critical temperature, as well as the appearance of the region of values $0.7 \leqslant n \leqslant 1$ in which charge ordering does not occur.

\bigskip
\bigskip

Figure~\ref{fig:afm-phasediag} shows a comparison of the values of $T_{AFM}$ obtained from the solution of equation \eqref{eq:TAFM} (dotted line) and in the MC method by the threshold value 0.01 of the antiferromagnetic order parameter (solid line). 
In addition to the characteristic decrease in critical temperature values, numerical simulations show the instability of the homogeneous antiferromagnetic (AFM) phase and the existence of a phase separation region (PS).
As is well known, considering phase separation leads to a significant modification of the phase diagrams of two-dimensional strongly correlated systems with AFM order, which is a complex problem \cite{Igoshev2010,Igoshev2015}. 
The ability to determine the boundaries of the PS region is a distinctive feature of the used MC method: certain values of n are never achievable for all given values of chemical potential $\mu$, since after annealing the system appears in the state corresponding to one of the boundary phases, in this case AFM and NO. 
The values of the order parameter are obtained as a linear interpolation of the boundary values, which corresponds to Maxwell's construction.

\bigskip
\bigskip

For the superfluid phase with the only non-zero parameters $\Delta=0.1$ and $t_b=1$, numerical simulations show that the phase is stable, and the largest difference between the critical temperature and the analytical value obtained from equation (9) is observed near $n=0$ (see Fig.\textbackslash{}eqref\{eq:TBS\}). 
For the P phase, the results are shown in Fig.~\ref{fig:fl-phasediag}. 
Numerical simulations show that the P phase is unstable at the chosen model parameters in a sufficiently wide region near $n=0$.

\bigskip
\bigskip
\bigskip

In summary, the results of numerical simulations within the modified MC method show good qualitative agreement with the results of the mean-field approximation, and the observed difference in the results is explained by taking into account the fluctuations of molecular fields in the heat bath algorithm. This allows us to use this numerical technique to analyze the properties of phase states in more complicated situations when an analytical solution is not available.

\bigskip
\bigskip

The research was supported by the Russian Science Foundation, grant no. 24-21-20147.


\end{document}